 \documentclass[final,3p,times]{elsarticle}

\usepackage{amssymb}
\usepackage{graphicx}
\usepackage{booktabs}

\journal{Physica A}

\begin{document}

\begin{frontmatter}


\title{Detecting the optimal number of communities in complex networks }
\author[]{Zhifang Li}
\author[]{Yanqing Hu\footnote{yanqing.hu.sc@gmail.com}}
\author[]{Beishan Xu}
\author[]{Zengru Di}
\author[]{Ying Fan\footnote{yfan@bnu.edu.cn}}
\address{Department of Systems Science, School of Management and
Center for Complexity
 Research, Beijing Normal University, Beijing 100875, China\\}

\begin{abstract}
To obtain the optimal number of communities is an important problem
in detecting community structure. In this paper, we extend the
measurement of community detecting algorithms to find the optimal
community number. Based on the normalized mutual information index,
which has been used as a measure for similarity of communities, a
statistic $\Omega(c)$ is proposed to detect the optimal number of
communities. In general, when $\Omega(c)$ reaches its local maximum,
especially the first one, the corresponding number of communities
\emph{c} is likely to be optimal in community detection. Moreover,
the statistic $\Omega(c)$ can also measure the significance of
community structures in complex networks, which has been paid more
attention recently. Numerical and empirical results show that the
index $\Omega(c)$ is effective in both artificial and real world
networks.
\end{abstract}


\end{frontmatter}

\section{Introduction}
Community detection has become a very important part of researches
on complex networks \cite{RMP47,PRE026121,PR175}. Communities or
modules mean high concentrations of edges within special groups of
vertices, and low concentrations between these groups
\cite{SIAM167,PNAS2658,PRE066111,PNAS8577}. Such communities have
been observed in many different fields. For instance, community
structure is a typical feature in social networks, where some of the
individuals can be part of a tightly connected group, others can be
completely isolated, while some others may act as bridges between
groups. Tightly connected groups of nodes in the World Wide Web
often correspond to pages on common topics, while communities in
genetic networks are related to functional modules. Consequently,
finding the communities within a network is a powerful tool for
understanding the structure and the function of the network, and its
growth mechanisms\cite{PR175}.

The general aim of community detection is to find meaningful
divisions into groups by investigating the structural properties of
the whole graph. There are two major aspects about this problem. One
concerns with proposing effective algorithms for detecting the
communities, and the other concerns with the significance or
robustness of the obtained divisions\cite{PR175,PR75}. For the first
aspect, many efficient heuristic methods have been proposed over the
years to detect the communities in networks, in particular those
based on spectral methods, divisive algorithms, modularity-based
methods, dynamic algorithms, and many others. However, most existing
algorithms are not able to get the optimal number directly.
Communities are just the final product of the algorithm. The
community number of each run may be different. For some previous
algorithms which can get the "optimal" community number depend on
modularity Q, including Greedy techniques, Simulated annealing,
Extremal optimization, Spectral optimization etc. By assumption,
high values of modularity indicate good partitions. So, the
partition corresponding to maximum value on a given graph should be
the best, or at least a good one. This is the main motivation for
modularity maximization, perhaps the most popular class of methods
to detect communities in graphs. However, exhaustive optimization of
Q is impossible, due to the maximum is out of reach, as it has been
proved that modularity optimization is an NP-complete problem
\cite{URL001907}, so it is probably impossible to find the solution
in a time growing polynomially with the size of the graph. Moreover,
Santo et al. have been strictly proved that even if modularity Q can
get the maximum value, the corresponding community number may not be
the optimal one\cite{PNAS36}. Most importantly, small changes of the
number of communities may have a huge impact on the results of
detecting the community structure. Therefore, it is essential to
look for simple and effective methods of detecting the optimal
number of communities.

   When we proceed the methods several times under the same condition,
they may give different community structures due to the random
factors in the algorithm. Then evaluating the quality of a partition
is also important in community identification. Newman described a
method to calculate the sensitivity of algorithms\cite{PRE066133}.
Danon et al. proposed a measurement based on information theory
\cite{JSMTE09008}. These two measurements mainly focus on the
proportion of nodes which are correctly grouped. Fan et al.
investigated the accuracy and precision of several algorithms
\cite{PhysicaA363}. Accuracy means the consistence when the
community structure from algorithm is compared with the presumed
communities, and precision is the consistence among the community
structures from different runs of an algorithm on the same network.
They proposed a similarity function \emph{S} to measure the
difference between partitions. The function \emph{S} integrates the
information about the proportion of nodes co-appearance in pair
groups of A, B and the total number of communities. Obviously, an
"ideal" community detection should be one that both with high
accuracy and high precision. In this paper, we propose a suitable
method for evaluating the optimal number of the communities based on
measuring the precision of algorithms and closely relating to the
accuracy of algorithms. We first use the algorithm based on mixture
model,which proposed by Newman and Leicht\cite{PNAS9564}, to induce
a sequence $ P(k) (1\leq{k}\leq{n})$ of divisions into communities;
Second, we measure the precision of the algorithm based on
"information entropy", which has been used to evaluate the
similarity of communities\cite{JSM10012,PRE066106,JMA873,PRE046119};
At last, we use our proposed index $\Omega(c)$ to find the optimal
number of the communities. Our statistic is an auxiliary method
which can be applied to almost all the algorithms with random
characteristics to help them find the optimal community number. It
is relatively simple to apply the method. It will not increase the
complexity of the algorithms, which just repeat the algorithm for
several times. A point we should mention is that our method needn't
to know the "real" community structures in advance.

This paper is organized as follows: The method is described in
detail in section 2. In section 3, we apply the method to several
artificial and real networks, and find some interesting results. In
section 4, some conclusions are given.

\section{Method}
First, for given number of groups \emph{c}, we use the method based
on mixture model to divide a network \emph{n} times, then we can get
\emph{n} (in this paper, $\emph{n}=50$) divisions of the same
network into communities. When we proceed the method several times
under the same condition, they may give different community
structures due to the random factors of the algorithm. So these
divisions generally have different community structures, meaning
that the communities in different divisions may include different
nodes and edges.

Second, we measure the precision of the algorithm based on comparing
the similarity between the divisions. A number of indexes for
measuring similarities or differences between partitions of a
network have been proposed in the past
\cite{PhysicaA363,JSM10012,PRE066106,JMA873,PRE046119,PRL078701,PNAS11433,PRE056135}.
Our method here follows the information theoretic methods. As
described in Ref.\cite{JSMTE09008,JSM10012},  a confusion matrix
\emph{N} was defined, where the rows correspond to the "real"
communities in networks, and the columns correspond to the "found"
communities. The element $N(i,j)$ is the number of nodes in the real
community \emph{i} that appear in the found community \emph{j}.
Therefore a measure of similarity between the partitions A and B is
\begin{equation}
R(c)=I{(A,B)}=\frac{-2\sum_{i=1}^{c_A}\sum_{j=1}^{c_B}N_{ij}log{(\frac{N_{ij}N}{N_{i.}N{.j}})}}
{\sum_{i=1}^{c_A}N_{i.}log{(\frac{N_{i.}}{N})}+\sum_{j=1}^{c_B}N_{.j}log{(\frac{N_{.j}}{N})}}
\end{equation}
Where A is the "real" community structure of the network, B denotes
the divisions of the network, and $c_A$, $c_B$ denote the numbers of
"real" communities and "found" communities respectively.
\emph{I(A,B)} is to measure the accuracy of the algorithm. The
larger \emph{I(A,B)} is, the better the community structure from
algorithm is consistent with the "real" one. We assume $c_A=c_B=c$,
then $I{(A,B)}$ can be simplified to
\begin{equation}
I_c{(a,b)}=\frac{-2\sum_{i=1}^c\sum_{j=1}^cN_{ij}log{(\frac{N_{ij}N}{N_{i.}N{.j}})}}
{\sum_{i=1}^cN_{i.}log{(\frac{N_{i.}}{N})}+\sum_{j=1}^cN_{.j}log{(\frac{N_{.j}}{N})}}
\end{equation}
    Both A and B in Eq.(2) are the divisions from the different runs of an
algorithm on the same network. Then $I_c{(a,b)}$ could measure the
precision of the algorithm. Here we use Eq.(2) to compare every two
different divisions of the same network into communities. Then for
each given \emph{c} and \emph{n}, we will get
$\frac{\emph{n}(\emph{n}-1)}{2}$ values of $I_c{(a,b)}$.

    Then, we propose an index $\Omega(c)$ as following,
\begin{equation}
\Omega(c)=\frac{\sum_{a=1,b=1,a<b}^{n}I_c{(a,b)}}{\frac{n(n-1)}{2}}
=\frac{2}{n(n-1)}\sum_{a=1,b=1,a<b}^{n}I_c{(a,b)}
\end{equation}
 $\Omega(c)$, actually, can also measure the precision of the
 algorithms.

  To show the effectiveness of our index $\Omega(c)$, we compare
it with $R(c)$ in a variety of networks whose community structures
are known. In general, the number of communities is much smaller
than the number of nodes in a network. Then, when the number of
communities is far fewer than the number of nodes, we find that
$\Omega(c)$ performs as well as $R(c)$ in measuring the similarities
between community structures. And we find that when it appears local
maximum, especially the first maximum, the corresponding \emph{c} is
likely the optimal number of groups.

\section{Results}
In order to investigate the performance of our index $\Omega(c)$, we
compare these two indices in ad hoc networks and some real networks
which have "known" community structures. To further measure the
performance of $\Omega(c)$, we apply it in several artificial
hierarchical networks which have "unknown" community structures in
advance, and we compare it with function \emph{Q} in ER random
networks. Finally, we intend to use $\Omega(c)$ to measure the
significance of community structure.

\textbf{3.1 Results of Binary ad hoc networks}

As a first test, we applied $\Omega(c)$ to computer-generated random
graphs with a well-known predetermined community
structure\cite{PRE026113}. Each graph has $N=128$ nodes divided into
$4$ communities of $32$ nodes each. Edges between two nodes are
introduced with different probabilities depending on whether the two
nodes belong to the same group or not: every node has $k_{intra}$
links on average to its neighbors in the same community and
$k_{inter}$ links to the outer world, keeping $<k_{intra}>$ +
$<k_{inter}>$ =16. In Fig.1, we show the performs of $\Omega(c)$ and
$R(c)$ in several Binary ad hoc networks. The curves correspond to
different choices for different $<k_{intra}>$. As we can see from
Fig.1, $\Omega(c)$ performances as well as $R(c)$, and when
$\Omega(c)$ reaches the first maximum value, the corresponding
number of groups $ c=4 $ is the optimal number of communities.

\begin{figure}
  \center
  \includegraphics[width=8cm]{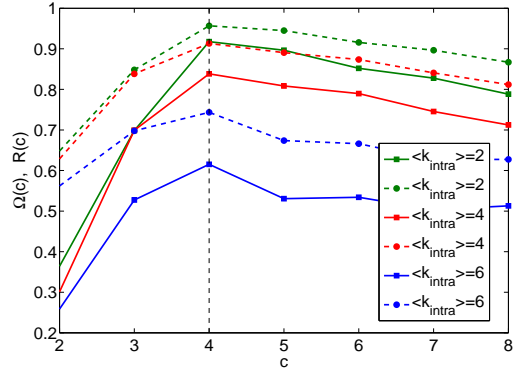}\\
  \caption{Comparing $\Omega(c)$ (shown with real lines) and $R(c)$
(shown with broken lines) in ad hoc networks with $n=128$, and 4
communities of 32 nodes each. The X axis denotes the number of
communities. The plot shows the results for three different numbers
of $<k_{intra}>$ corresponding to 2, 4, or 6. Both $\Omega(c)$ and
$R(c)$ achieve their first maximum when the number of communities is
4. The result is averaged by 100 times.}
\end{figure}

\textbf{3.2 Results on Zachary's karate club network}

The Zachary karate club network has been considered as a simple
sample for community-detecting
methodologies\cite{PNAS9564,PRE026113,PRE016115,JGAA191,PRE027104,PRE046108,EPL18009}.
This network is constructed with the data collected from observing $
34 $ members of a karate club over a period of 2 years and
considering friendship between members. It has been proved that the
best partition of this network has two communities by many previous
algorithms\cite{PNAS8577,PNAS9564,JSM10012,PRE026113}. We also apply
$\Omega(c)$ and $R(c)$ to the Zachary karate club network. As shown
in Fig.2, $\Omega(c)$ appears its first maximum value when c=2.
According to our method, $c=2$ is the optimal number of communities
in the Zachary karate club network, which correspond to the results
given in\cite{PNAS8577,PNAS9564}.

\begin{figure}
  \center
  \includegraphics[width=8cm]{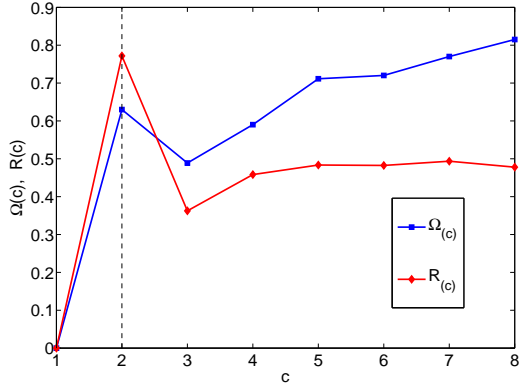}\\
  \caption{Comparing $\Omega(c)$ (shown with square line) and
$R(c)$ (shown with diamond line) in the Zachary's karate club
network. Both of them achieve the first maximum when the number of
communities is 2. The result is averaged by 100 times.}
\end{figure}

\textbf{3.3 Results on an LFR Benchmark}

As we have remarked above, all vertices of ad hoc networks have
approximately the same degree and all communities have exactly the
same size by construction. These two features are at odds with what
is observed in graph representations of real systems. Degree
distributions of real networks are usually skewed, with many
vertices with low degree coexisting with a few vertices with high
degree. Lancichinetti, Fortunato, and Radicchi considered that a
good benchmark should not be assumed that all communities have the
same size: the distribution of community sizes of real networks is
also broad, with a tail that can be fairly well approximated by a
power law. They introduced a class of benchmark graphs which account
for the heterogeneity in the distributions of both degree and
community size\cite{PRE046110,PRE016118}. Such benchmark is a more
faithful approximation of real-world networks with community
structure than simpler benchmarks like, e. g. that by Girvan and
Newman. Through the method in\cite{PRE046110}, we get a class of LFR
Benchmark networks and we apply index $\Omega(c)$ to one of them.
The result is shown in Fig.3. As we can see from the figure, when
$c=10$, $\Omega(c)$ reaches the first maximum value, which
corresponds to the condition of the network we construct. In fact,
$\Omega(c)$ relates to the number of communities. When the community
number c is large enough, $\Omega(c)$ is bound to increase with the
increase of c. When the community number is equal to the network
size, $\Omega(c)$ will be one. Our statistic is generally effective
on the networks which their community number is much smaller than
the vertex number, which is consistent with the characteristic of
most real networks.

\begin{figure}
  \center
  \includegraphics[width=8cm]{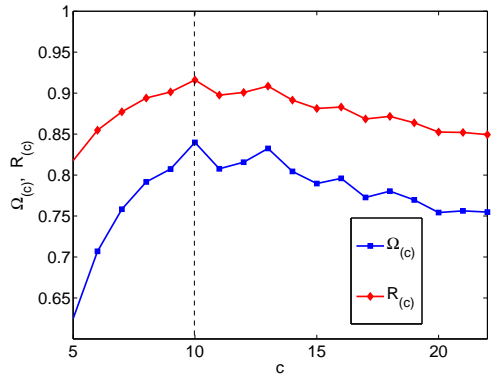}\\
  \caption{Comparing $\Omega(c)$ (shown with square line) and $R(c)$
(shown with diamond line) in one of LFR benchmark networks, which
has a skewed degree distribution, similar to real networks. The
number of nodes N=500; Each node is given a degree taken from a
power law distribution with exponent \emph{$\gamma$}=3; The average
node degree $<k>$=20; Each node shares a fraction $1-\mu$ of its
links with the other nodes of its community and a fraction $\mu$
with the other nodes of the network; $\mu$=0.2 is the mixing
parameter; The sizes of the communities are taken from a power law
distribution with exponent $\beta$=2. Through the method, we get the
network and the number of communities $c$=10. Both $\Omega(c)$ and
$R(c)$ appear the first local maximum value when the number of
communities is 10. }
\end{figure}

\textbf{3.4 Results on hierarchical networks}

    Hierarchical networks have been mentioned in many literatures
\cite{PRE016110,Nature98,PNAS13773,PRE016127,PNAS15224,JSMTE04024}.
Marta Sales-Pardo et al. proposed a method to construct hierarchical
nested random graphs\cite{PNAS15224}. In this paper, we test our
method on hierarchical artificial networks with two levels. Taking
Fig.4(a) as an example, we explain how we construct the hierarchical
networks. We first create a network with 80 nodes that at the first
level has two modules comprising 40 nodes each. Once having assigned
nodes to groups, we draw an edge between a pair of nodes $(i, j)$
with probabilities

 (\emph{1.}) $p_2$($p_2=\frac{d_2}{c_2}$), where $d_2$ is the average
degree of the module at the second level and $c_2$ is the number of
nodes in the module), if $(i, j)$ are in the same module at the
second level;
 (\emph{2.}) $p_1$($p_1=\frac{d_1}{c_1}$), where $d_1$
is the average degree of the module at the first level and $c_1$ is
the number of nodes in the module), if $(i, j)$ are in the same
module at the first level;
 (\emph{3.}) $p_0$($p_0=\frac{d_0}{c_0}$),
where $d_0$ is the average degree of the module at the top level and
$c_0$ is the number of nodes in the whole network, otherwise. We
impose that $p_2 > p_1 > p_0$ (here $p_2=0.85$, $p_1=0.25$,
$p_0=0.01$), then the resulting network has a larger density of
connections between nodes grouped in the same submodule at the
second level, a smaller density of connections between groups of
nodes grouped in the same module at the first level, and an even
smaller density of connections between nodes grouped in separate
modules at the top level. Thus, the network has by construction an
artificial hierarchical organization. Fig. 4 shows the results of
$\Omega(c)$ in two of these kinds of networks. In Fig.4, $\Omega(c)$
appears several local maximum values, and we can find that the
corresponding numbers of communities of the first two maximum values
of $\Omega(c)$ are well corresponding to the number of communities
of different levels in hierarchical networks. However, $\Omega(c)$
can not always detect the "optimal" community number of hierarchical
networks effectively, but just with a certain probability. For
example, we have calculated the probabilities of the above two
hierarchical networks. For the 80-nodes network, the probabilities
of obtaining a maximum when c=2 and c=4 are 0.91, 0.32 respectively.
For the 180-nodes network, the probabilities of obtaining a maximum
when c=3 and c=9 are 0.85, 0.21 respectively. What's more, as shown
in fig.4(b), $\Omega(2) > \Omega(4)$; and as shown in fig.4(d),
$\Omega(3) > \Omega(9)$. This is a very interesting phenomenon, which
means large community structures are more likely to be identified by the algorithms.

\begin{figure}
\center
       \includegraphics[width=7cm]{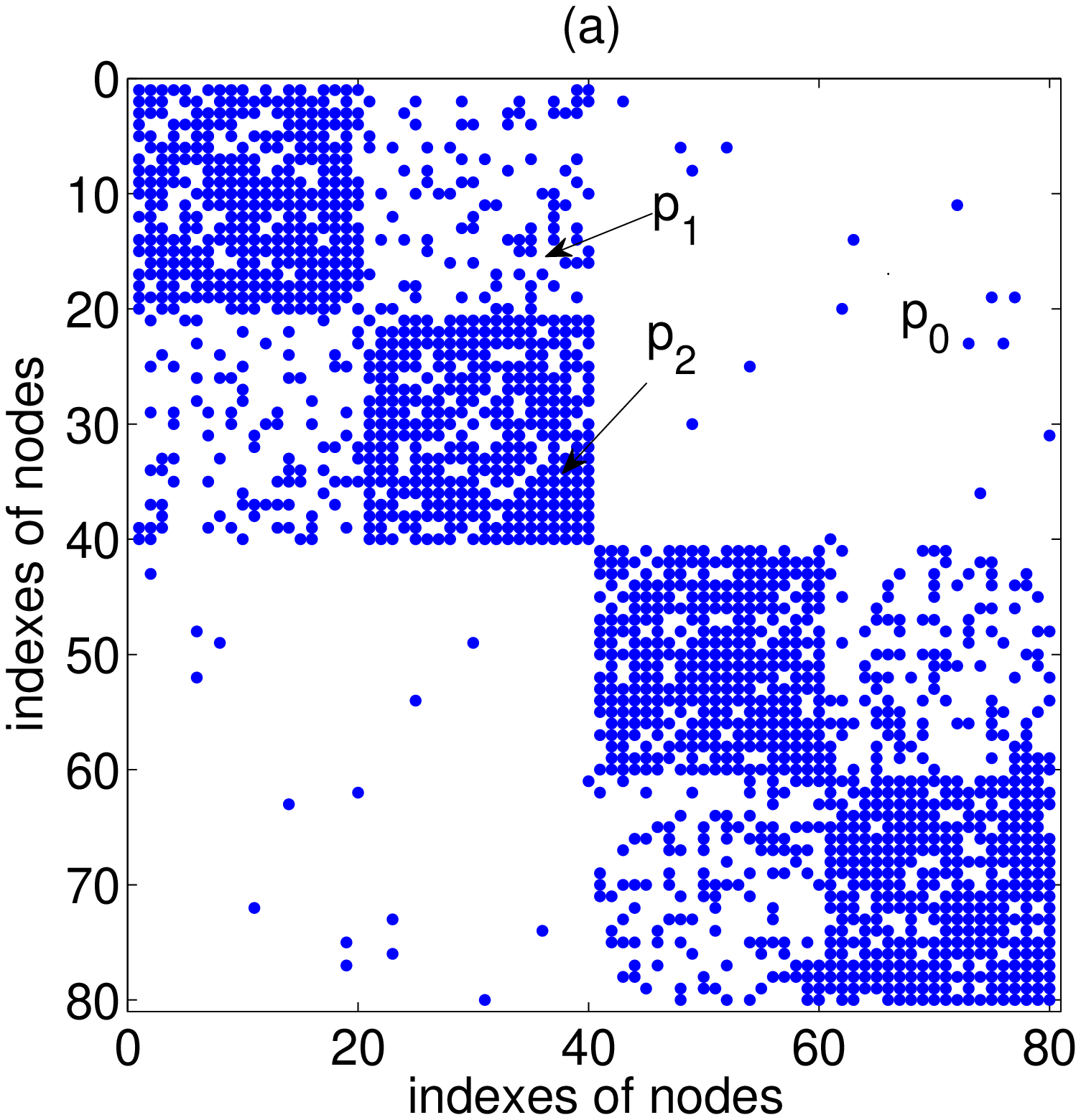}
        \includegraphics[width=7cm]{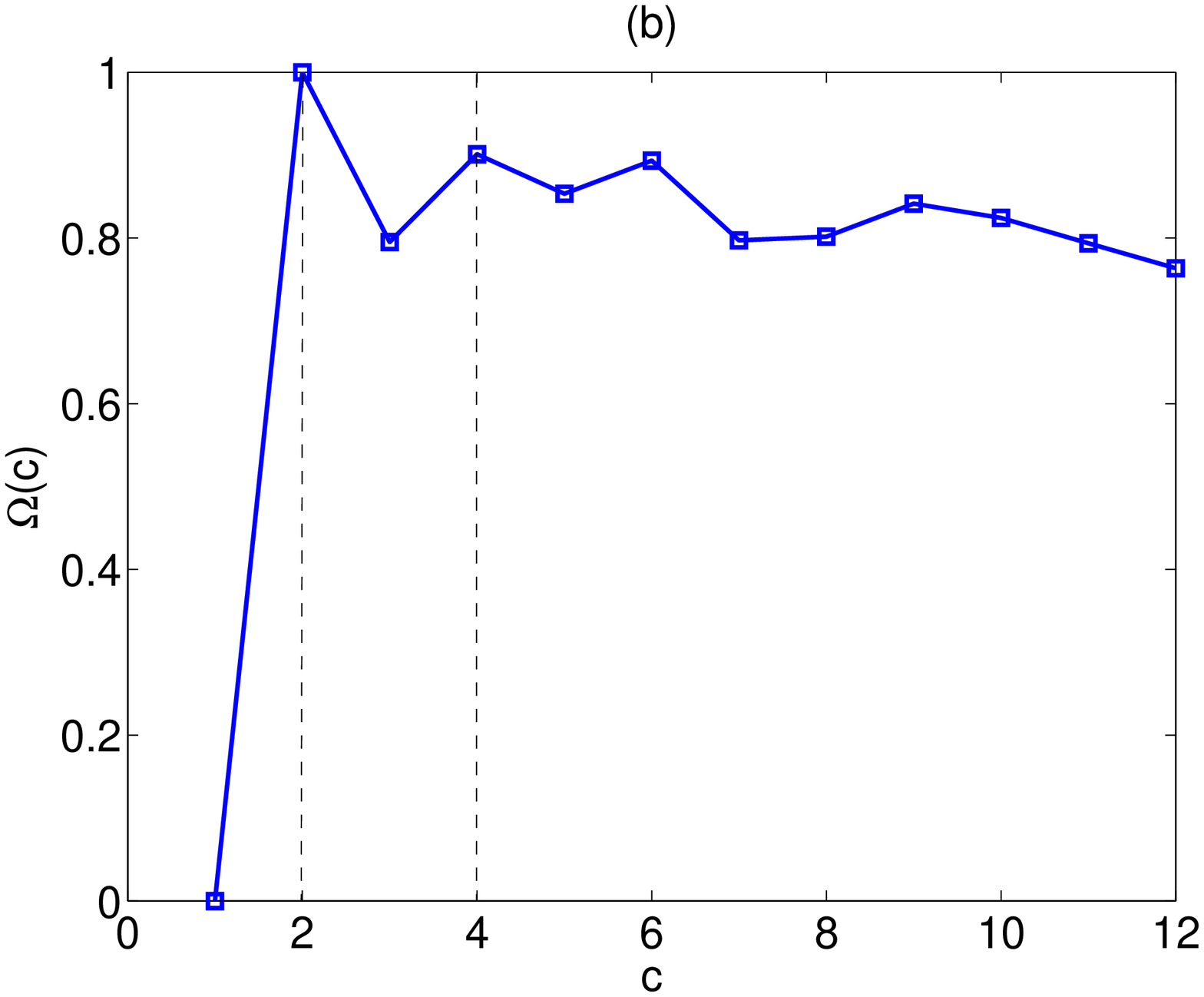}\\
        \includegraphics[width=7cm]{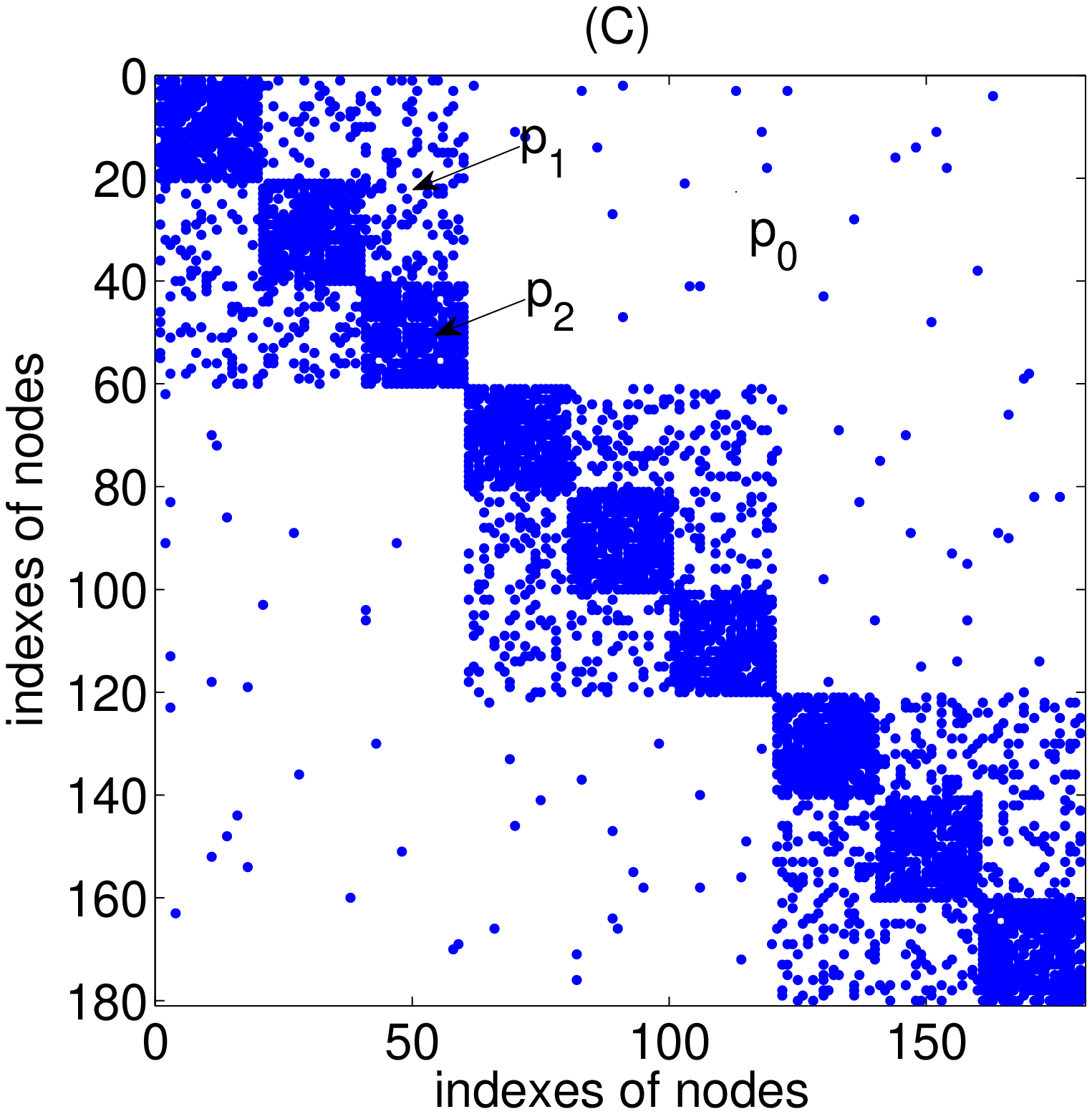}
        \includegraphics[width=7cm]{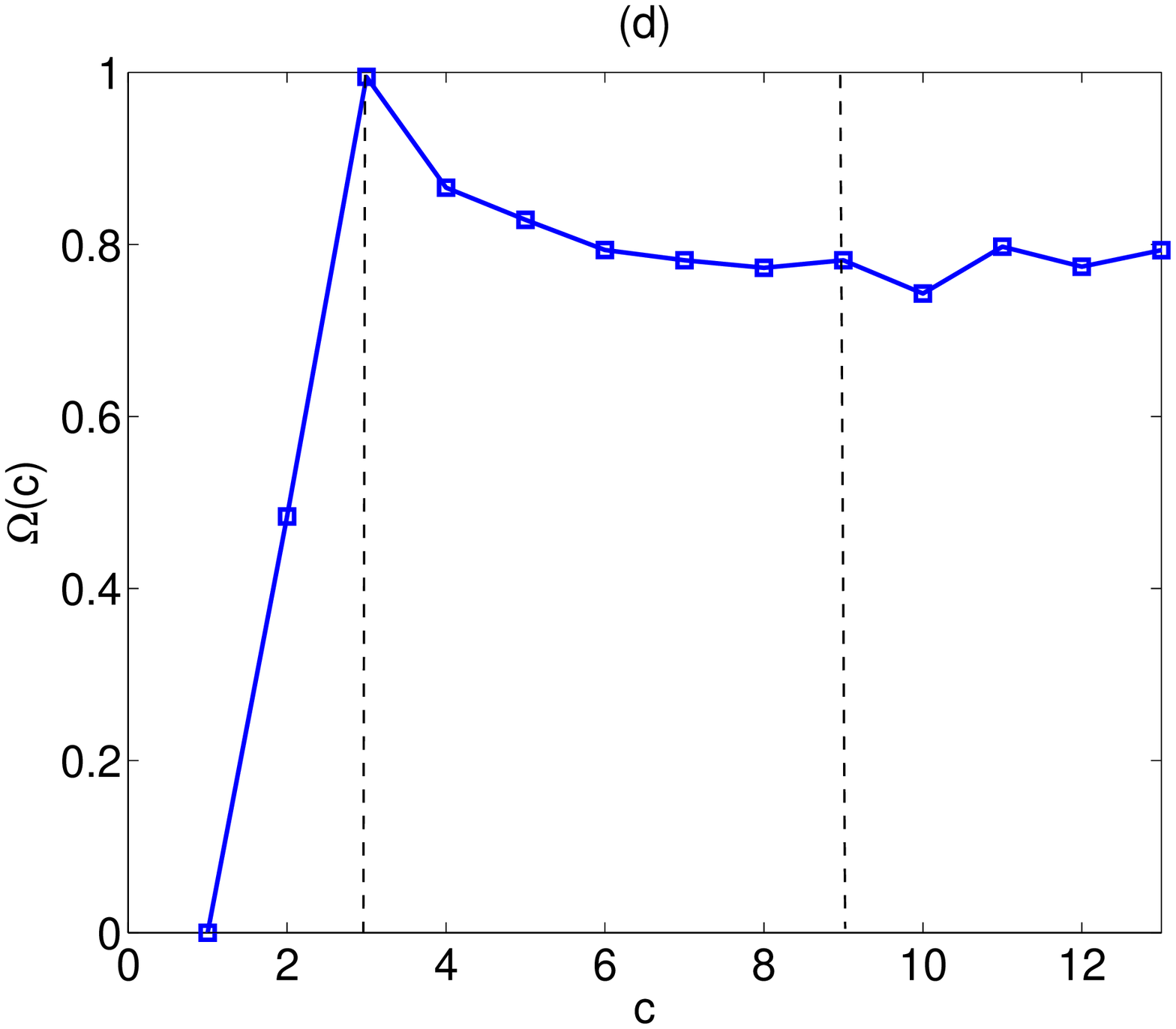}
\caption{The performance of $\Omega(c)$ in computer-generated
hierarchical networks. In (a), (c), both the x axis and the y axis
present the number of nodes $\emph{n}$ in the networks. In (b), (d),
the x axis presents the number of communities. From the plots we can
see that our method is also helpful for detecting the community
structures at different levels in artificial hierarchical networks.
(a) $n=80$, $p_0=0.002$, $p_1=0.1$, $p_2=0.85$; (c) $n=180$,
$p_0=0.005$, $p_1=0.2$, $p_2=0.8$;}
\end{figure}

\textbf{3.5 Results on an ER random network}

    Ref.\cite{PR75,PRE046110(9),PRE026102} show that high values of the modularity of Newman
and Girvan does not necessarily indicate that a graph has a definite
cluster structure. It in particular shows that partitions of random
graphs may also achieve considerably large values of \emph{Q},
although we do not expect them to have community structure, due to
the lack of correlations between the linking probabilities of the
vertices. We compare the index $\Omega(c)$ with the modularity
function \emph{Q} in an ER network, which have $128$ nodes and $<k>
= 1.5$. The network is normally considered with indefinite community
structure. We use the extremal optimization
algorithm\cite{PRE027104} to detect the community structure of the
network. It is divided into $14$ communities and the maximal
\emph{Q} is 0.6056, which is large enough to consider the network
has definite community structure. It means the modularity function
\emph{Q} doesn't performs well in networks which have indefinite
community structures.  We apply $\Omega(c)$ to the network. As shown
in Fig.5, the statistic $\Omega(c)$ doesn't appear maximums.
According to our method, it can't find the optimal number of
communities in the network, meaning that the network doesn't have
definite community structure. Thus $\Omega(c)$ may shed light on
evaluating whether a network has definite community structure or
not.

\begin{figure}
  \center
  \includegraphics[width=8cm]{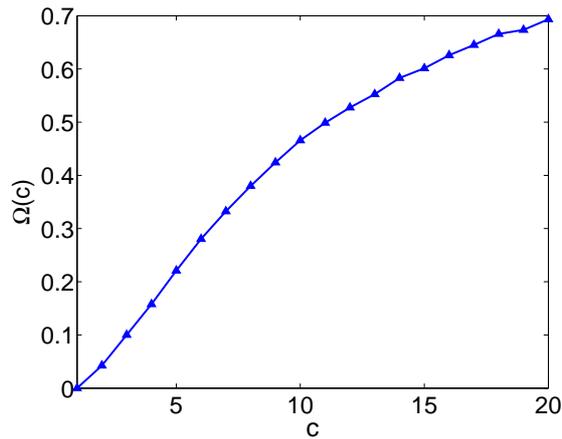}\\
  \caption{The performance of $\Omega(c)$ in an ER random
network, which has N=128 nodes and $<k> = 1.5$. This random network
is normally considered with no definite community structure.
However, the maximum of modularity function \emph{Q} calculated by
extremal optimization algorithm is 0.6056, which is large enough to
consider the network has definite community structure. $\Omega(c)$
doesn't appear maximums. It means that the network doesn't have
definite community structure, which correspond with the real
condition. The result is averaged by 100 times.}
\end{figure}

\textbf{3.6 Measuring the significance of community structures}

    Many efficient methods have been proposed for finding communities,
but few of them can evaluate the communities found are significant
or trivial definitely\cite{PRE066106}. In some works the concept of
significance has been related to that of robustness or stability of
a partition against random perturbations of the graph structure. The
basic idea is that, if a partition is significant, it will be
recovered even if the structure of the graph is modified, as long as
the modification is not too extensive. Instead, if a partition is
not significant, one expects that minimal modifications of the graph
will suffice to disrupt the partition, so other clusterings are
recovered\cite{PR75}. We apply the statistic $\Omega(c)$ to evaluate
the significance of communities in binary ad hoc networks. As
described in $3.1$, each of these computer-generated networks has
$N=128$ nodes divided into $4$ communities of $32$ nodes each and we
know the optimal number of communities of them is $4$. By treating
$c=4$, we apply $\Omega(4)$ and $R(4)$ to measure the significance
of these networks. With $<k_{intra}>$ increasing, the community
structure of the network will become less and less significative. As
shown in Fig.6, both $\Omega(4)$ and $R(4)$ are descending with the
increase of $<k_{intra}>$, which indicates $\Omega(c)$ performs well
in measuring the significance of community structures in complex
networks.

\begin{figure}
  \center
  \includegraphics[width=8cm]{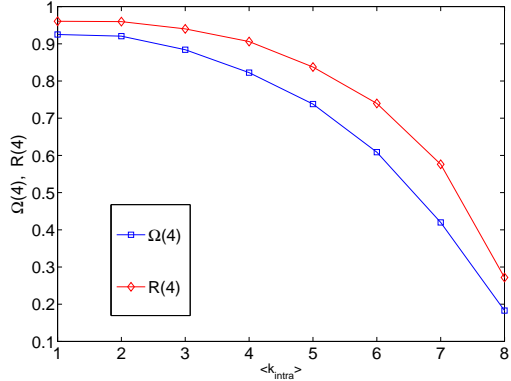}\\
  \caption{The performance of $\Omega(c)$ (shown with blue square line) and $R(c)$
(shown with red diamond line) in measuring the significance of
community structures in computer-generated networks. Both
$\Omega(c)$ and $R(c)$ are descending with the increase of
$<k_{intra}>$ due to the community structures becoming fuzzier and
fuzzier. The X axis gives the change of $<k_{intra}>$. The result is
averaged by 100 times.}
\end{figure}

\section{Conclusion and Discussion}
The investigation of the optimal number of communities in a network
is an important and tough issue in the study of complex networks. In
this paper, we present a method to detect the optimal number of
communities in complex networks based on the information theoretic
ideas. We apply the index $\Omega(c)$ to some networks,including
artificial networks and real networks with well-known community
structures. The results show that when the number of communities is
much smaller than the number of nodes, the index is effective on
normal networks, LFR benchmark networks and hierarchical community
structure networks. For hierarchical networks, $\Omega(c)$ can just
detect the "optimal" community number of hierarchical networks with
a certain probability and we have found a very interesting
phenomenon, which large community structures are more likely to be
identified by the algorithms. This phenomenons will be further
considered in our future work. Moreover, the index $\Omega(c)$ can
be used to measure the significance of community structure which has
been paid much attention recently. The statistic can nearly work
based on every community detecting algorithm with the character of
randomization, and it will not change the complexity of the
algorithm.

\section*{Acknowledgement}
The authors wish to thank Professor Shlomo Havlin for many
helpful suggestions. This work is partially supported by
NCET-09-0228 and NSFC under the grants No. 60974084 and 70771011. Y.
Hu was supported by Scientific Research Foundation and Excellent
Ph.D Project of Beijing Normal University.

\end{document}